# Robustness of the Digital Image Watermarking Techniques against Brightness and Rotation Attack

Harsh K Verma[1], Abhishek Narain Singh[2], Raman Kumar[3]

[1,2,3]Department of Computer Science and Engineering
Dr B R Ambedkar National Institute of Technology
Jalandhar, Punjab, India.
E-mail: [1]vermah@nitj.ac.in, [2]singhabhi444@gmail.com, [3]er.ramankumar@aol.in

*Abstract-* **The recent advent in the field of multimedia proposed a many facilities in transport, transmission and manipulation of data. Along with this advancement of facilities there are larger threats in authentication of data, its licensed use and protection against illegal use of data. A lot of digital image watermarking techniques have been designed and implemented to stop the illegal use of the digital multimedia images. This paper compares the robustness of three different watermarking schemes against brightness and rotation attacks. The robustness of the watermarked images has been verified on the parameters of PSNR (Peak Signal to Noise Ratio), RMSE (Root Mean Square Error) and MAE (Mean Absolute Error).**

*Keywords-* **Watermarking, Spread Spectrum, Fingerprinting, Copyright Protection.**

## I. INTRODUCTION

Advancements in the field of computer and technology have given many facilities and advantages to the human. Now it becomes very easier to search and develop any digital content on the internet. Digital distribution of multimedia information allows the introduction of flexible, cost-effective business models that are advantageous for commerce transactions. On the other hand, its digital nature also allows individuals to manipulate, duplicate or access media information beyond the terms and conditions agreed upon. Multimedia data such as photos, video or audio clips, printed documents can carry hidden information or may have been manipulated so that one is not sure of the exact data. To deal with the problem of trustworthiness of data, authentication techniques are being developed to verify the information integrity, the alleged source of data, and the reality of data [1]. Cryptography and steganography have been used throughout history as means to add secrecy to communication during times of war and peace [2].

### A. Digital Watermarking

Digital watermarking involves embedding a structure in a host signal to "mark" its ownership [3]. We call these structures digital watermarks. Digital watermarks may be comprised of copyright or authentication codes, or a legend essential for signal interpretation. The existence of these watermarks within a multimedia signal goes unnoticed except when passed through an appropriate detector. Common types of signals to watermark are still images, audio, and digital video. To be effective a watermark must be [4]:

- *Unobstructive*; that is, it should be unperceivable when embedded in the host signal.
- *Discreet*; unauthorized watermark extraction or detection must be arduous as the mark's exact location and amplitude are unknown to unauthorized individuals.
- *Easily extracted*; authorized watermark extraction from the watermarked signal must be reliable and convenient.
- *Robust/fragile to incidental and unintentional distortions*; depending on the intended application, the watermark must either remain intact or be easily modified in the face of signal distortions such as filtering, compression, cropping and re-sampling performed on the watermarked data.

In order to protect ownership or copyright of digital media data, such as image, video and audio, encryption and watermarking techniques are generally used. Encryption techniques can be used to protect digital data during the transmission from sender to the receiver. Watermarking technique is one of the solutions for the copyright protection and they can also be used for fingerprinting, copy protection, broadcast monitoring, data authentication, indexing, medical safety and data hiding [5].

## II. WATERMARK EMBEDDING AND EXTRAXTION

A watermark, which is often consists of a binary data sequence, is inserted into a host signal with the use of a key [6]. The information embedding routine imposes small signal changes, determined by the key and the watermark, to generate the watermarked signal.

This embedding procedure (Fig. 1) involves imperceptibly modifying a hoist signal to reflect the information content in the watermark so that the changes can be later observed with the use of the key to ascertain the embedded bit sequence. The process is called watermark extraction.

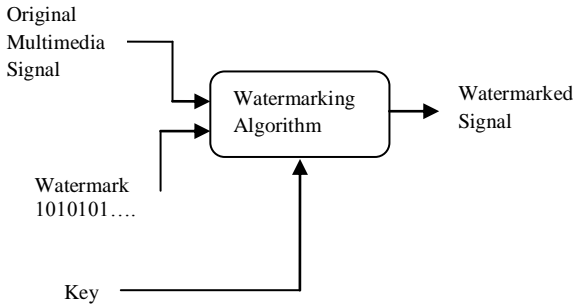

**Fig. 1** Watermark embedding process

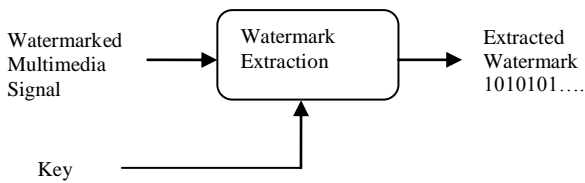

**Fig. 2** Watermark extraction process

The principal design challenge is in embedding the watermark so that it reliably fulfills its intended task. For copy protection applications, the watermark must be recoverable (Fig. 2) even when the signal undergoes a reasonable level of distortion, and for tamper assessment applications, the watermark must effectively characterize the signal distortions. The security of the system comes from the uncertainty of the key. Without access to this information, the watermark cannot be extracted or be effectively removed or forged.

### III. WATERMARKING TECHNIQUES

Three different watermarking techniques each from different domain i.e. Spatial Domain, Frequency Domain and Wavelet Domain [7] watermarking have been chosen for the experiment. The techniques used for the comparative analysis of watermarking process are CDMA Spread Spectrum watermarking in spatial domain, Comparison of mid band DCT coefficients in frequency domain and CDMA Spread Spectrum watermarking in wavelet domain [8].

### A. CDMA Spread Spectrum Watermarking in Spatial Domain
The algorithm of the above method is given below:

1. To embed the watermark

   a. Convert the original image in vectors
   b. Set the gain factor k for embedding
   c. Read in the watermark message and reshape it into a vector
   d. For each value of the watermark, a PN sequence is generated using an independent seed
   e. Scatter each of the bits randomly throughout the cover image
   f. When watermark contains a '0', add PN sequence with gain k to cover image
      i. if  watermark(bit) = 0
          watermarked_image=watermarked_image + k*pn_sequence
      ii. Else if  watermark (bit) = 1
          watermarked_image=watermarked_image + pn_sequence
   g. Process the same step for complete watermark vector

2. To recover the watermark

   a. Convert back the watermarked image to vectors
   b. Each seed is used to generate its PN sequence
   c. Each sequence is then correlated with the entire image
      i. If ( the correlation is high)
          that bit in the watermark is set to "1"
      ii. Else
          that bit in the watermark is set to "0"
   d. Process the same step for complete watermarked vector
   e. Reshape the watermark vector and display recovered watermark

### B. Comparison of Mid-Band DCT Coefficients in Frequency Domain

The algorithm of the above method is given below:

3. To embed the watermark

   a. Process the image in blocks.
   b. For each block
      Transform block using DCT.
      If  message_bit is 0.
          If dct_block $(5,2) < (4,3)$ .
          Swap them.
      Else If $(5, 2) > (4,3)$
          Swap them.
      If $(5, 2) - (4,3) < k$
          $(5, 2) = (5, 2) + k/2;$
          $(4, 3) = (4, 3) - k/2;$
      Else
          $(5, 2) = (5, 2) - k/2;$
          $(4, 3) = (4, 3) + k/2;$
   c. Move to next block.

4. To recover the watermark

   a. Process the image in blocks.

b. For each block

      Transform block using DCT.
          If (5,2) > (4,3)
             Message = 1;
          Else
             Message=0;

c. Process next block.

*C. CDMA Spread Spectrum Watermarking in Wavelet Domain*

The algorithm of the above method is given below:

5. To embed the watermark

    a. Convert the original image in vectors
    b. Set the gain factor k for embedding
    c. Read in the watermark message and reshape it into a vector
    d. Do Discrete Wavelet Transformation of the cover image
        i. [cA,cH,cV,cD]=dwt2(X,'wname')computes the approximation coefficients matrix cA and details coefficients matrices cH, cV, and cD (horizontal, vertical, and diagonal, respectively), obtained by wavelet decomposition of the input matrix X. The 'wname' string contains the wavelet name
    e. Add PN sequence to H and V components
        ii. If (watermark == 0)
             cH1=cH1+k*pn_sequence_h;
             cV1=cV1+k*pn_sequence_v;
    f. Perform Inverse Discrete Wavelet Transformation
        iii. watermarked_image = idwt2(cA1,cH1,cV1,cD1,'wname',[Mc,Nc])

6. To recover the watermark

    a. Convert back the watermarked image to vectors
    b. Convert the watermark to corresponding vectors
    c. Initialize watermark vectors to all ones
        i. Watermark_vector = ones (1,$M_W$*$N_W$)
          where, $M_W$= Height of watermark.
          $N_W$ = Width of watermark.
    d. Find correlation in H and V components of watermarked image
        i. correlation_h()=corr2(cH1,pn_sequence_h);
        ii. correlation_v()=corr2(cV1,pn_sequence_v);
        iii. correlation(wtrmkd_img)=(correlation_h() + correlation_v())/2;
    e. Compare the correlation with mean correlation
        i. if (correlation(bit) > mean(correlation))
             watermark_vector(bit)=0;
    f. Revert back the watermar_vector to watermark_image

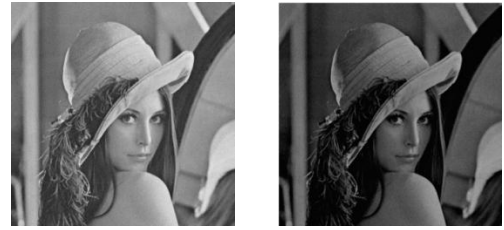

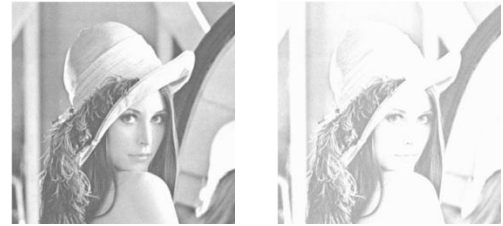

**Fig. 3** Brightness Attack (a) Original watermarked image, (b) Watermarked image after -25% brightness, (c) Watermarked image after 25% brightness, (d) Watermarked image after 50% brightness

## IV. BRIGHTNESS ATTACK

Brightness attack is one of the most common types of attack on digital multimedia images. Three different levels of brightness attacks have been done. First the brightness is increased by -25% i.e. decreased by 25%, again the brightness is increased by 25% and at last the brightness is increased by 50%. The brightness attack has been shown below in Fig. 3.

Table 1 shows the results of brightness attack on different watermarking techniques on various parameters like Peak Signal to Noise Ratio [9], average Root Mean Square Error and average Mean Absolute Error.

## V. ROTATION ATTACK

Rotation attack is among the most popular kinds of geometrical attack on digital multimedia images [10]. Three levels of rotations have been implemented. First the original watermarked image is being rotated by 90 degree, then 180 degree and at last the image is being rotated by 270degree in clock wise direction. The rotation attack has been shown below in Fig. 4.

The results of the rotation attack have been shown below in Table 2 for all the three watermarking schemes.

## VI. EXPERIMENTAL RESULTS

The comparative analysis of the three watermarking schemes has been done on the basis of brightness and rotation attacks. Results of the individual watermarking technique have been compared on the basis of PSNR (Peak Signal to Noise Ratio), RMSE (Root Mean Square Error) and MAE (Mean Absolute Error).

**Table 1** Performance analysis of watermarking techniques against Brightness Attack

| | | PSNR (dB) | Avg. RMSE | Avg. MAE |
|---|---|---|---|---|
| -25% | CDMA SS in Spatial D. | 19.982 | 25.55 | 2.563 |
| | Comp. of Mid Band DCT | 22.427 | 19.283 | 1.459 |
| | CDMA SS in Wavelet D. | 26.215 | 12.466 | 0.61 |
| 25% | CDMA SS in Spatial D. | 17.203 | 35.184 | 4.86 |
| | Comp. of Mid Band DCT | 21.339 | 21.855 | 1.874 |
| | CDMA SS in Wavelet D. | 24.305 | 15.533 | 0.946 |
| 50% | CDMA SS in Spatial D. | 16.921 | 36.345 | 5.185 |
| | Comp. of Mid Band DCT | 18.435 | 30.534 | 3.659 |
| | CDMA SS in Wavelet D. | 22.605 | 18.892 | 1.40 |

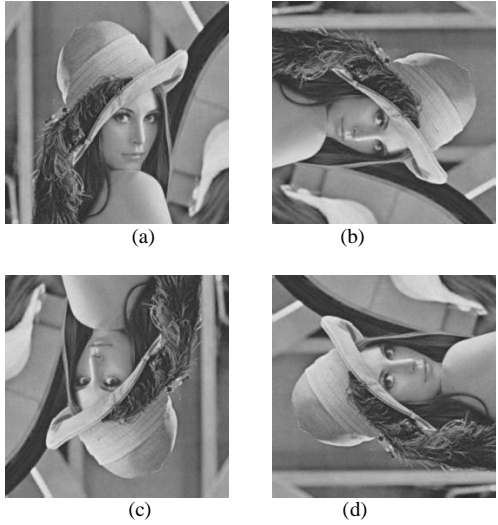

**Fig. 4** Rotation Attack (a) Original watermarked image, (b) Watermarked image after 90 degree rotation, (c) Watermarked image after 180 degree rotation, (d) Watermarked image after 270 degree rotation

**Table 2** Performance analysis of watermarking techniques against Rotation Attack

| | | PSNR (dB) | Avg. RMSE | Avg. MAE |
|---|---|---|---|---|
| 90° | CDMA SS in Spatial D. | 12.286 | 61.973 | 15.087 |
| | Comp. of Mid Band DCT | 15.602 | 42.305 | 7.035 |
| | CDMA SS in Wavelet D. | 19.50 | 27.008 | 2.866 |
| 180° | CDMA SS in Spatial D. | 11.193 | 70.283 | 19.398 |
| | Comp. of Mid Band DCT | 11.999 | 64.059 | 16.122 |
| | CDMA SS in Wavelet D. | 13.517 | 53.784 | 11.368 |
| 270° | CDMA SS in Spatial D. | 11.818 | 65.404 | 16.803 |
| | Comp. of Mid Band DCT | 13.533 | 53.685 | 11.323 |
| | CDMA SS in Wavelet D. | 16.251 | 39.263 | 6.057 |

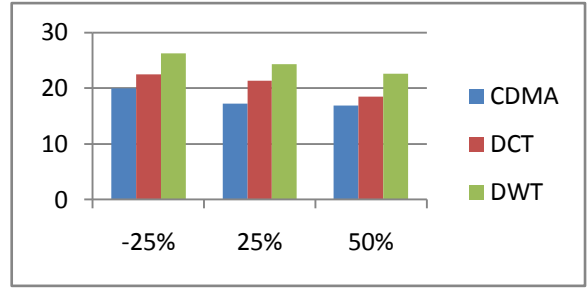

**Fig. 5** Graph showing PSNR values for brightness attack on different watermarking schemes

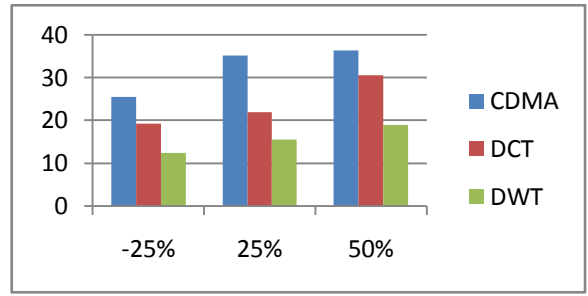

**Fig. 6** Graph showing Average RMSE values for brightness attack on different watermarking schemes

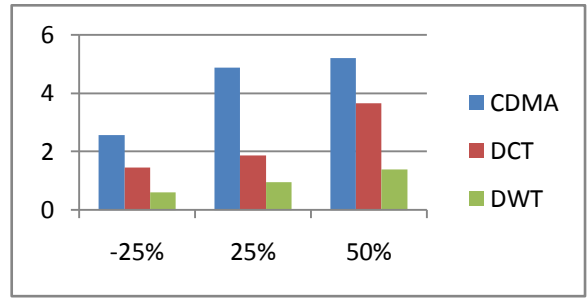

**Fig. 7** Graph showing Average MAE values for brightness attack on different watermarking schemes

### A. Results of Brightness Attack

The Fig. 5, 6 and 7 show the results of brightness attack on all the three techniques of watermarking. A comparative analysis is done thereafter.

The graphs shown above are the comparative results of the brightness attack on the three watermarking techniques discussed. Greater the PSNR value implies more robust is the watermarking technique against attack. Having a look on the Fig. 5 the DWT (CDMA SS watermarking in Wavelet domain) technique is proved to be the best candidate for the digital image watermarking, since its having greater PSNR value than the other two techniques. Similarly from Fig. 6 and 7, the values of Root Mean Square Error and Mean Absolute Error are also minimum for the Discrete Wavelet Transform

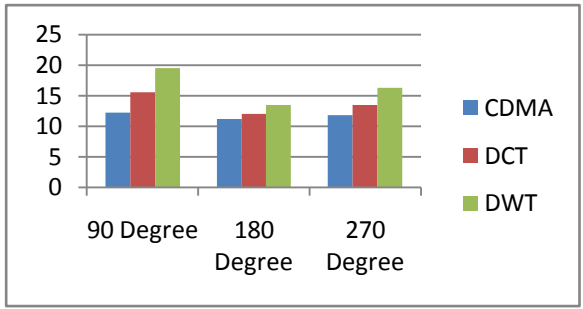

**Fig. 8** Graph showing PSNR values for rotation attack on different watermarking schemes

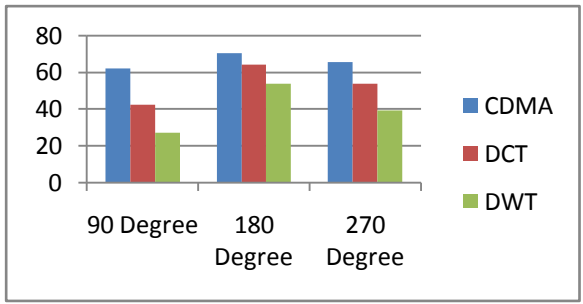

**Fig. 9** Graph showing Average RMSE values for rotation attack on different watermarking schemes

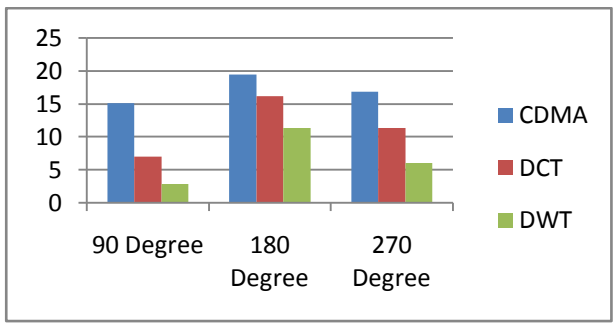

**Fig. 10** Graph showing Average MAE values for rotation attack on different watermarking schemes

domain technique, so it is being proved to be the best against brightness attack.

### B. Results of Rotation Attack

A comparison of the results of rotation attack is being done by showing the results in the form of Fig. 8, 9 and 10. Graphs are drawn for all the three parameters of evaluation. A comparative analysis of the result has been done thereafter.

The PSNR values in Fig. 8 shows that the CDMA SS watermarking in Wavelet domain technique is having the greatest value for the PSNR value. This shows that the wavelet domain watermarking is the best practice for the digital image watermarking purpose.

Experimental values of the brightness and rotation attack shows that the CDMA Spread Spectrum watermarking technique is the best choice for the watermarking of digital multimedia images.

Discrete Cosine Transformation Domain shows somewhat greater robustness against the rotation attack. Spatial domain watermarking techniques are not good candidates for large size of watermarks. They show poor results with larger size of watermarks.

### VII. CONCLUSIONS

This paper focuses on the robustness of the watermarking techniques chosen from all the three domains of watermarking against brightness and rotation attack. The key conclusion of the paper is that the Wavelet domain watermarking technique is the best and most robust scheme for the watermarking of digital multimedia images. This work could further be extended to the watermarking purpose of another digital content like audio and video.